\documentstyle[psfig]{l-aa} 

\def\fdg{\hbox{$.\!\!^\circ$}}          
\def\etal{{et\,al. }}

\begin{document}
 
   \thesaurus{02               
              (11.17.3         
               13.07.1)        
             }

   \title{Gamma-Ray Bursts from Radio-Quiet Quasars}
 
   \author{N. Schartel\inst{1,2}, H. Andernach\inst{3,4}, J. Greiner\inst{5}} 
 
   \offprints{N. Schartel}
 
   \institute{$^1$ ESA, IUE Observatory,
                   P.O. Box 50727, E-28080 Madrid, Spain \\
              $^2$ Affiliated to the
                   Astrophysics Division, Space Science Department, ESTEC, \\
              $^3$ INSA; ESA IUE Observatory,
                   P.O. Box 50727, E-28080 Madrid, Spain \\
              $^4$ Depto.~de Astronom\'\i a, Univ.~Guanajuato,
                   Apdo.~Postal 144, 36000 Guanajuato, Mexico \\
              $^5$ Max-Planck-Institut f\"ur extraterrestrische Physik,
                   85740 Garching, Germany\\
              }

   \date{Received July 17, 1996; accepted }
 
   \maketitle
  
\begin{abstract}
We study positional coincidences bet\-ween gamma-ray bursts 
in the BATSE 3B catalogue and  quasars/AGN taken from the 
V\'eron-Cetty \& V\'eron compilation.
For most classes of AGN, for BL Lac objects and for radio-loud 
quasars, we find no excess coincidences above random expectation,
and we give upper limits for their burst emission rate.

However, surprising evidence is found for a positional correlation
between gamma-ray bursts and radio-quiet quasars.
A total of 134 selected bursts with a position error radius $<$1\fdg8 and the 
nearest and intrinsically brightest radio-quiet quasars 
($M_v < -24.2$ and $z < 1.0$) have a probability of $>$99.7\% 
to be associated with each other.  
An analysis of a smaller sample of well-localized interplanetary network
gamma-ray burst positions supports this result.
This correlation strongly favours the cosmological origin of gamma-ray bursts
and enables to estimate its distance scale.
The average luminosity of those gamma-ray bursts which we associate directly 
with radio-quiet quasars is of the order of $4\times10^{52}$ erg
(assuming isotropic emission).


\end{abstract}

\section{Introduction}

The nature of gamma-ray bursts (GRBs) remains an enigma, even nearly 30 years 
after their discovery with the Vela satellites in 1967 (Klebesadel et al. 1973).
The bursts detected with the
Burst and Transient Source Experiment (BATSE) on board the {\it
Compton Gamma-Ray Observatory} (CGRO) show that their positions
are completely consistent with isotropy. 
For the  subsample of 1005 GRBs in the 3B BATSE catalogue the 
Galactic dipole moment is within $0.9\,\sigma$ of the value expected 
for isotropy, and the observed Galactic quadrupole moment 
deviates only $0.3\,\sigma$ (Briggs et al.\ 1996) from isotropy.

Thus, GRBs are distributed significantly more isotropic than any known
Galactic population and most of the Galactic models are no longer 
in agreement with the data (cf.\ e.g.\ Briggs et al. 1996, Hartmann 
et al. 1995). 
A spherical dark matter halo model could still be consistent 
with the measured burst positions, but requires a core radius 
larger than  necessary to explain the Galaxy's 
rotation curve (Hakkila et al.\ 1994; Caldwell \& Ostriker 1981; 
Briggs et al.\ 1996). Extended halo models have to place the typical BATSE 
GRB farther away than about 100 kpc in order to obtain the measured isotropy. 
On the other hand, the observed bursts must be closer than 
about 350 kpc, because no excess of bursts is observed in the direction of M31.
Neutron stars born with velocities exceeding the local Galactic escape velocity 
have been proposed to match these requirements (Podsiadlowski \etal 1995), 
but in order to obtain isotropy of bursts, only a very small 
fraction of low-velocity neutron stars are allowed to burst in this model.
Moreover, a delayed turn-on is necessary to avoid a concentration of bursts 
to the Galactic disk (see e.g.\ Li \& Dermer 1992; Briggs et al.\ 1996).

The alternative possibility, a burst location at cosmological distances, 
does not require to postulate additional restrictions (Paczy\'nski 1986). Also,
the observed inhomogeneity of the bursts' radial distribution
can be regarded as supportive evidence for this idea.
The number of faint bursts 
shows a deficit with respect to the power law of slope $-3/2$ 
of a homogeneous distribution in Euclidean space (Schmidt et al.\ 1988).
This can be explained naturally if faint bursts 
are located at high redshift (Paczy\'nski 1991; Dermer 1992; Piran 1992).
There have been many attempts to estimate the redshift range of GRB
sources from their brightness distribution (e.g. Fenimore \etal 1993b; 
Fenimore \& Blom 1995).
Whereas a redshift of about 0.1 for the nearest bursts seems 
likely, only a very crude estimate of the redshift of the faintest 
bursts can be made ($z = 0.8$ through $ z = 3.0$).
Apart from the uncertainty in  the cosmological parameters,
such studies suffer from the unknown evolution function of the burst
sources and the corresponding  emission rate (see Sect. \,4.1)
as  function of redshift (Rutledge et al.\ 1995).

The realization that gamma-ray measurements alone may be not sufficient
to solve the GRB mystery has prompted increasing attempts for counterpart
searches of GRBs at various wavelengths.
Two different methods  are possible to search  for GRB  counterparts:

\noindent{\bf (1)}
Because of the typically large errors of burst positions and the huge number
of quiescent objects within each GRB error box, 
one method consists of simultaneous observations at other
wavelengths (X-ray, optical, IR, radio) to search for transient sources
during the GRB event (see recent reviews by Hartmann 1995, Greiner 1995). 
As a result of these studies, there is solid evidence that gamma-ray bursts 
do not produce simultaneous optical flashes brighter than about 5$^{\rm th}$ 
mag for an assumed duration of  1 sec (Greiner \etal 1996) or 
simultaneous radio emission stronger than 10$^{-13}$ erg cm$^{-2}$ s$^{-1}$
(Cortiglioni \etal 1981; Inzani \etal 1982).

\noindent{\bf (2)} 
The second, more statistical approach is the search for positional 
coincidences of GRB locations with comparison samples of any other known 
class of astronomical object. 
Various types of catalogued objects, both Galactic and extragalactic,
were compared with the BATSE gamma-ray error boxes, but any 
correlation between these and the bursts was compatible with 
random expectation.
(i) Webber et al.\ (1995) looked for catalogued counterparts of
60 bursts with the smallest error boxes available ($\sim0.25\, {\rm deg}^2$)
and found no position  coincidences above random expectation for 
extragalactic objects (see, however, Sect. 5.2.2. for our investigation of 
the latter sample).
(ii) Gorosabel \etal (1995) found no significant correlation between 44 GRBs 
localized by the WATCH instrument to less than one square degree and objects 
(galactic and extragalactic) listed in 33 different catalogues.
(iii) Extremely rare events in external galaxies, for example merging neutron
stars, are one possibility to explain an extragalactic origin of GRBs
(Eichler \etal 1989; Paczy\'nski 1991).
Under such a hypothesis a correlation between burst positions and  
the large-scale structure of matter should be expected. 
The concentration of nearby  galaxies towards the
super-galactic plane offers the possibility to search for
counterparts up to a redshift of $z \sim 0.1$.
Hartmann et al.\ (1996) found no compelling evidence for a concentration of
GRBs to the super-galactic plane in the 3B catalogue.
(iv) With  a likelihood method that compares the counts-in-cell
distribution of gamma-ray bursts with the one expected 
from the known large-scale structure of the Universe,
Quashnock (1996) argued that the nearest bursts originate at
a redshift $z\sim0.03$ and the farthest ones at $z > 0.25$.
(v) However, searches for a correlation between GRB and the  
direction of rich clusters of galaxies (situated right in the
above redshift range) did not arrive at consistent results
(Howard et al.\ 1993, Nemiroff \etal 1994, Deng et al.\ 1995,
Kolatt \& Piran 1996, Marani \etal 1996). From a  comparison of the 
angular correlation functions of GRBs
and that of extragalactic radio sources, Kurt \& Komberg (1995) 
concluded that gamma-ray bursts do not belong to a normal population
of galaxies with $z<1.5$. 

If we assume that GRBs are emitted at cosmological
distances and keep in mind that no clear correlation was found with  either
clusters of galaxies or   the super-galactic plane
then it seems unlikely that any normal constituents of galaxies 
can cause gamma-ray bursts. The most promising  sources
with a different nature are quasars and active galactic nuclei (AGN). 
They had been suggested as GRB sources by Prilutski \& Usov (1975)
shortly after the discovery of GRBs (Klebesadel \etal 1973).

Both, the number of bursts with relatively small error boxes, and
the number of identified AGN and quasars are increasing permanently. 
Webber et al.\ (1995) used the 1989 version 
of the V\'eron-Cetty \& V\'eron compilation of QSOs and AGN.
The fact that the number of known quasars and AGN has doubled since 
then has motivated us to check their results with  improved  statistics. 
A further chance to find correlations arises if
only a certain subclass of the AGN or quasars are
counterparts of GRB. In that case the mean distance between the members of 
the sample will increase and the error of the gamma-ray burst position
will play a less important role. We have therefore not only repeated the 
work of Webber \etal (1995) with larger samples, but extended the search for
correlations to various subclasses of AGN.

Our paper is arranged as follows. In the first part of Sect. 2 the
GRB sample and the methods  used to search for correlations
are introduced. In the second part  we provide the 
samples of AGNs and quasars as well as different  subclasses.
Our search for correlations between GRBs and various subclasses is given in 
the first part of Sect. 3. In the second part we study 
in more detail those subclasses 
which show a positive correlation. The  final results are 
described in Sect. 4 and we elaborate on the one positive correlation found.
Finally, we discuss our results in the light of previous 
studies in Sect. 5 and conclude with some remarks in Sect. 6.

\section{The Samples and the Analysis Method}

\subsection{The sample of observed bursts}

The 3B catalogue of GRBs (Meegan \etal 1996) provides the position of 1122 
events, together with a statistical error, $r_{st}$, which is defined by the 
formal covariance matrix from  a $\chi^2$-test on the assumption of normal 
errors. The circle with this radius centered on the fitted position of the 
burst corresponds to the 68\% confidence ellipse.
This error is based on the Poisson uncertainty in the BATSE measurement
of the burst flux  by each Large Area Detector and is believed to have
a Gaussian distribution (cf.\ the description of the 3B catalogue).
For many  bursts the additional systematic error of 1\fdg6 is larger 
than the formal error.
Its distribution is not very well known, but may have a more extended
tail than a Gaussian (Pendleton et al. 1996). 
Combination of both errors results in a total error, $  R_t$, of
\begin{equation}
R_t = \sqrt{ \left(  r_{st} \right)^2 + \left(  1\fdg6 \right)^2 } 
\end{equation}
which is an estimate of the 68\% confidence interval for the burst location.

In order to test a given source sample of QSO or AGN ({\it ``comparison
sample''} in what follows)
for correlation with GRBs we calculate the number of coincidences. 
A comparison with the number of correlations obtained for simulated burst
positions allows us to estimate the significance of the correlations obtained
for the real burst positions. To test for any bias introduced by either 
the distribution of GRB positions or the non-homogeneous distribution 
of the comparison sample we will use three different
methods. They differ in the way of calculating the number of coincidences
and in the way the simulated burst positions are generated and are described 
in the next paragraphs.

Due to the uneven exposure of the sky the burst distribution is not isotropic.
However, since the exposure map is primarily symmetric with respect to the 
ecliptic (except the south atlantic anomaly), and the quasar distribution 
is asymmetric with respect to the galactic plane (due to the search techniques),
the affect on the coincidence statistics is negligible.

\subsection{Randomly simulated GRB samples}

We calculate the number of GRBs containing at least one object of a given 
comparison sample within their 68\% confidence interval.  
To reduce the noise all bursts which have 
a positional distance $ d \le  R_t$ to a source in the comparison sample
are counted only once.

Until now there are no reports on the presence of any large-scale structure
(like e.g.\ dipole or higher moments) in the GRB distribution.
Therefore it appears reasonable to use randomly simulated control samples 
of GRBs to obtain the number of chance coincidences, and their
distribution.  However, we should keep in mind that
there is no ultimate proof for the assumption that randomly distributed 
positions reproduce the characteristics of the observed GRB distribution.
To estimate the significance of the correlation we simulate 1000 different 
sets of randomly distributed GRBs and determine the corresponding number of 
coincidences, $C$.
The number of coincidences expected by chance, $<C_B>$, is given 
by the mean over the 1000 values of $C$.

\begin{figure}[thbp]
\psfig{figure=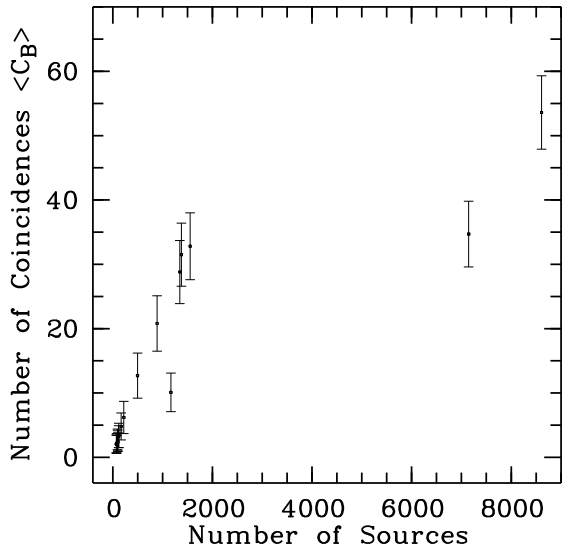,width=7.6cm
       ,bbllx=2.7cm,bblly=4.7cm,bburx=13.5cm,bbury=15.1cm,clip=}
      \vspace{-0.35cm}
      \caption{The expected number of position coincidences between
               Gamma-ray bursts and sources are plotted
               against the number of sources in the studied sample.
               The used burst sample is given through a maximal
               position uncertainty of $R_t <$ 1\fdg8 and contains
               136 gamma-ray bursts. If GRBs and sources of the comparison
               sample were distributed isotropically on the sky, the expected 
               number of position coincidences would asymptotically approach 
               136 at more than $\approx$4000 sources. However, quasars in the
               V\'eron-Cetty \& V\'eron catalog are distributed very irregularly
               with nearly no occurrences in the galactic plane while 
               clustering in fields where deep searches have been performed
               (especially for high-redshift quasars).}
\vspace{-0.3cm}
\end{figure}

Further, the one thousand values of C represent the
distribution for coincidences between the comparison sample
and simulated burst sample which (by definition) do not show any
correlation (null-hypothesis). This distribution will be used
to apply statistics on the coincidences found for the real samples.

Since every burst is counted only once, the maximum possible number of
coincidences is equal to the number of GRBs used.
This  method is justified since any observed burst can have only
one counterpart.
With this method of counting only once, the obtained number of coincidences
depends both on the number of objects in the comparison sample as well
as on their distribution on the sky, in particular their isotropy.
In Fig. 1 the value $<C_B>$ is plotted against the size of the
various comparison samples of QSOs/AGNs we studied (see Sect. 3 for details).
Clearly, $C_B$ is not a linear function of the size of the
comparison sample.
A linear correlation holds for most of the samples
with less than approximately 2000 members. For larger samples the
$<C_B>$ values obtained are significantly lower than expected from a
linear extrapolation of the smaller samples. However, with the 1000
simulations  of the GRB positions we establish
the distribution of the coincidences for a given sample.
This enables us to obtain the corresponding probability for
a real correlation of GRBs with the respective comparison sample.

\subsection{Shifted GRB sample}

The small-scale structure in the GRB locations is more problematic 
than the large-scale structure.  Although no evidence for an autocorrelation 
in the 3B catalogue has been reported, a possible intrinsic correlation of GRB
positions can not be excluded at present, given their large positional 
uncertainties. Rather than testing the 3B catalogue itself for the presence of 
small-scale structure, we will test our results using a method 
intrinsically robust against a possible hidden small-scale structure in
the GRB distribution.

We applied the same method as in the previous section to calculate the number 
of coincidences. However, a different method to construct the
simulated random positions was introduced to determine the
number of chance coincidences.  One thousand control
samples were generated by changing the sign of the declination
of the measured burst positions and shifting the right ascension
1000 times with a random number in the range between $5\degr$ and
$255\degr$. The random samples constructed in such a way conserve the
small-scale structure of the real GRB distribution and allow to test
its influence on the correlations.

\subsection{Singular Coincidences}

By using  the {\it shifted GRB sample} we are able to test our results against 
a possible small-scale structure of the GRB position distribution.
Because the QSO/AGN positions provided in the V\'eron-Cetty \& V\'eron 
(1996) compilation are not distributed homogeneously over the sky, we 
need a method which is robust against ``clustering'' of objects in
the comparison samples. This is especially important as most of the 
``clustering'' is due to the accumulation of objects in 
several deep small-area searches for AGN and QSOs.
To this end we counted only bursts with a single QSO or AGN in their 
1\,$\sigma$ error circle. With this method we exclude every burst 
with two or more QSO/AGN counterparts, i.e.\ too many to allow a 
direct comparison with the large error boxes of GRBs.
The expected background values and
the corresponding expected distribution for an uncorrelated
burst sample were obtained again with {\it randomly simulated GRB samples}.

The statistical errors of the GRB positions
range from 0\fdg13 to 34\fdg15. Including the systematic
error this results in a range from 1\fdg61 to 34\fdg18 for the
total error.   The mean total error for all bursts is 5\fdg2$\pm$4\fdg1,
and the mean area of the error circles is 138 deg$^{\rm 2}$ per burst.
Therefore, the sum of all 68\% confidence error boxes of 3B bursts 
corresponds to more than three times the whole sky.  In order to avoid too many
spurious coincidences, we restricted our study to the bursts with
small errors.  In Table 1 we list the characteristics of the different
GRB samples used here. As mentioned above, both the statistical and total errors
depend on the Poisson uncertainty in the BATSE measurements.
Therefore, bursts with higher flux generally show smaller positional errors.
In Fig. 2 the total error radii for 835 bursts are shown versus
the burst flux in the 20--50\,keV energy band.
Thus, restricting the bursts to those with a smaller error circle
also restricts them to a higher flux range.
If we search for positional coincidences of GRBs with AGNs and quasars
and thus assume that GRBs originate at cosmological distances,
then the restricted flux range may correspond to a restricted redshift range.

\begin{table}[thp]
  \vspace{-0.35cm}
  \caption{Characteristics of the GRB Samples Used}
\begin{tabular}{ccr} 
\hline \hline
\noalign{\smallskip}
\multicolumn{2}{c}{Maximal Error} & Number \\
   total & ~~statistical~~ & of bursts \\
\noalign{\smallskip}
\hline 
\noalign{\smallskip}
$R_t <$ 1\fdg7 &$ r_{st} <$ 0\fdg574 &   80    \\
$R_t <$ 1\fdg8 &$ r_{st} <$ 0\fdg825 &  134    \\
$R_t <$ 2\fdg0 &$ r_{st} <$ 1\fdg200 &  206    \\
\noalign{\smallskip}
\hline
\end{tabular}
  \vspace{-0.35cm}
\end{table}

\subsection{The AGN/Quasar Samples Studied}

We used the latest (7$^{\rm th}$) version of the ``Catalogue of
Quasars and Active Nuclei'' (V\'eron-Cetty \& V\'eron 1996).
This catalogue has been compiled from the available literature
and provides the position of 11,662 sources drawn from 1662 references.
Due to several deep small-area searches for AGN and QSOs the objects
in this compilation are not distributed homogeneously over the sky,
but appear ``clustered'' in these areas. 
Therefore, any test for a positional correlation between GRBs and quasars 
or AGN must be robust against observational bias inherent in the analyzed 
sample.

\begin{figure}[phb]
\psfig{figure=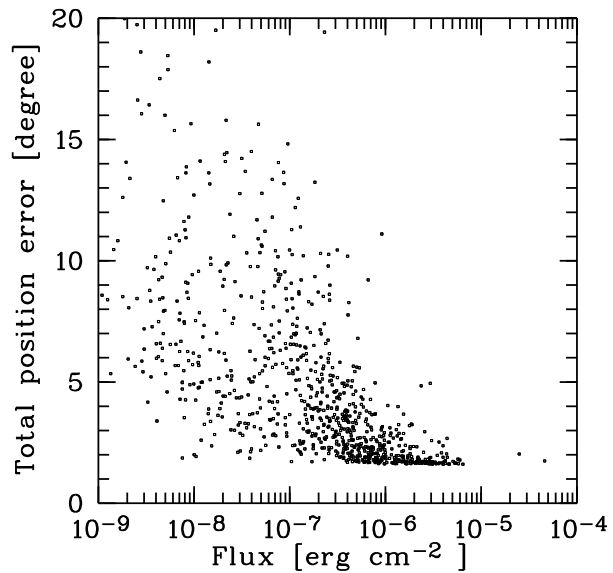,width=8.2cm
       ,bbllx=2.6cm,bblly=4.5cm,bburx=14.0cm,bbury=15.4cm,clip=}
      \vspace{-0.2cm}
      \caption{The total position error is
               shown over the measured  flux in the 20--50 keV band
              for 835  bursts. The total position error is always
              greater than the systematic error of 1\fdg6.}
\end{figure}

\begin{table*}[thbp]
  \caption{Comparison samples of QSO/AGN and positional coincidences with 
three GRB samples of different positional errors. $<C_B>$ is the expected 
number of chance coincidences derived from 1000 randomly simulated GRB samples.
${C_{R}}_{t}$ gives the number of coincidences with the 3B catalog bursts and 
P(\%) is the fractional number of simulations revealing less coincidences than 
${C_{R}}_{t}$.}
\begin{tabular}{llrrrrrrrrr} \hline \hline
\noalign{\medskip}
\multicolumn{1}{c}{Main}
&  \multicolumn{1}{c}{Subclass} & N$^{(a)}$
 & \multicolumn{1}{c}{$\bar{D}^{(b)}$}& \multicolumn{2}{c}{$R_t$=1\fdg7}
& \multicolumn{2}{c}{$R_t$=1\fdg8}
& \multicolumn{2}{c}{$R_t$=2\fdg0}\\
\multicolumn{1}{c}{Class} & 
\multicolumn{1}{c}{or} &  & \multicolumn{1}{c}{(\degr)} \\
& \multicolumn{1}{c}{selection}    &   & \multicolumn{1}{c}{} &
$<C_B>$ & \multicolumn{1}{c}{$C_{1\fdg7}/P(\%)$} &
$<C_B>$ & \multicolumn{1}{c}{$C_{1\fdg8}/P(\%)$} &
$<C_B>$ & \multicolumn{1}{c}{$C_{2\fdg0}/P(\%)$} \\ 
\noalign{\medskip}
\hline 
\noalign{\smallskip}
QSO & all            &$8609$&$ 0.6\pm0.9 $&$ 31.3\pm4.2 $&$ 36 $ / $ 83.4  $ &
                                   $ 53.6\pm5.7 $&$ 63 $ / $ 94.2  $ &
                                   $ 85.3\pm7.0 $&$ 98 $ / $ 95.2  $  \\
QSO & radio-quiet    &$7146$&$ 0.4\pm0.9 $&$ 20.3\pm3.9 $&$ 28 $ / $ 96.4  $ &
                                   $ 34.7\pm5.1 $&$ 44 $ / $ 94.6  $ &
                                   $ 55.3\pm6.4 $&$ 66 $ / $ 94.0  $  \\
QSO & radio-loud     &$1377$&$ 2.3\pm1.6 $&$ 18.2\pm3.6 $&$ 15 $ / $ 16.3  $ & 
                                   $ 31.5\pm4.9 $&$ 35 $ / $ 73.1  $ &
                                   $ 51.1\pm6.2 $&$ 58 $ / $ 84.8  $  \\
QSO & highly polarized &$  72$&$ 10.\pm7.4 $&$  1.2\pm1.1 $&$  1 $ / $ 31.5  $ &
                                   $  2.0\pm1.4 $&$  3 $ / $ 67.8  $ &
                                   $  3.4\pm1.7 $&$  5 $ / $ 74.8  $  \\
BL & all              &$220$&$ 5.9\pm4.3 $&$  3.5\pm1.9 $&$  2 $ / $ 13.5  $ &
                                   $  6.2\pm2.5 $&$  3 $ / $  ~4.7  $ &
                                   $ 10.3\pm3.2 $&$  5 $ / $  ~2.0  $  \\
BL & confirmed        &$ 93$&$ 8.6\pm5.2 $&$  1.6\pm1.2 $&$  2 $ / $ 52.9  $ &
                                   $  2.8\pm1.6 $&$  2 $ / $ 21.6  $ &
                                   $  4.5\pm2.1 $&$  3 $ / $ 16.4  $  \\
BL & highly polarized   &$ 76$&$ 11.\pm6.8 $&$  1.3\pm1.1 $&$  1 $ / $ 27.6  $ &
                                   $  2.2\pm1.5 $&$  2 $ / $ 35.1  $ &
                                   $  3.7\pm1.9 $&$  3 $ / $ 28.7  $  \\
BL & radio selected   &$119$&$ 8.4\pm5.5 $&$  1.9\pm1.4 $&$  0 $ / $  ~0.0  $ &
                                   $  3.4\pm1.9 $&$  1 $ / $  ~3.5  $ &
                                   $  5.7\pm2.4 $&$  2 $ / $  ~2.7  $  \\
BL & X-ray selected   &$ 82$&$ 8.7\pm5.0 $&$  1.4\pm1.2 $&$  2 $ / $ 59.5  $ &
                                   $  2.4\pm1.5 $&$  2 $ / $ 31.2  $ &
                                   $  4.0\pm2.0 $&$  3 $ / $ 24.0  $  \\
AGN & AGN                &$1553$&$ 2.1\pm1.7 $&$ 19.0\pm4.0 $&$ 20 $ / $ 55.1  $ &
                                   $ 32.8\pm5.2 $&$ 34 $ / $ 55.5  $ &
                                   $ 53.2\pm6.4 $&$ 55 $ / $ 57.4  $  \\
AGN & Seyfert 1          &$ 888$&$ 2.7\pm2.2 $&$ 12.1\pm3.4 $&$ 13 $ / $ 55.7  $ &
                                   $ 20.8\pm4.3 $&$ 21 $ / $ 47.7  $ &
                                   $ 34.0\pm5.4 $&$ 36 $ / $ 64.3  $  \\
AGN & Seyfert 2          &$ 496$&$ 3.9\pm2.6 $&$  7.3\pm2.6 $&$  6 $ / $ 26.2  $ &
                                   $ 12.7\pm3.5 $&$ 12 $ / $ 38.2  $ &
                                   $ 21.0\pm4.4 $&$ 20 $ / $ 38.2  $  \\
AGN & Prob. Seyfert   &$  97$&$ 8.0\pm6.1 $&$  1.4\pm1.2 $&$  3 $ / $ 82.5  $ &
                                   $  2.6\pm1.6 $&$  3 $ / $ 54.6  $ &
                                   $  4.3\pm2.1 $&$  3 $ / $ 19.5  $  \\
AGN & LINER              &$  71$&$ 10.\pm7.8 $&$  1.2\pm1.1 $&$  0 $ / $  ~0.0  $ &
                                   $  2.1\pm1.4 $&$  1 $ / $ 14.8  $ &
                                   $  3.4\pm1.9 $&$  2 $ / $ 14.1  $  \\
AGN &  radio-quiet       &$1346$&$ 2.2\pm1.8 $&$ 16.7\pm3.8 $&$ 18 $ / $ 58.6  $ &
                                   $ 28.8\pm4.9 $&$ 31 $ / $ 62.8  $ &
                                   $ 46.8\pm6.1 $&$ 50 $ / $ 66.9  $  \\
AGN & radio-loud         &$ 166$&$ 7.4\pm4.2 $&$  2.8\pm1.6 $&$  1 $ / $  ~6.7  $ &
                                   $  4.8\pm2.1 $&$  1 $ / $  ~0.4  $ &
                                   $  7.9\pm2.8 $&$  2 $ / $  ~0.3  $  \\ 
    & AG$^{(c)}$&$1165$&$ 1.0\pm2.1 $&$  5.8\pm2.3 $&$  9 $ / $ 87.5  $ &
                                   $ 10.1\pm3.0 $&$ 18 $ / $ 98.7  $ &
                                   $ 16.3\pm3.9 $&$ 25 $ / $ 98.1  $  \\
    & Nucl. H\,II &$ 116$&$ 7.7\pm5.9$ &$  1.8\pm1.3 $&$  1 $ / $ 15.7  $ &
                                   $  3.2\pm1.7 $&$  1 $ / $  ~3.6  $ &
                                   $  5.3\pm2.2 $&$  2 $ / $  ~2.6  $  \\
\noalign{\medskip}
\hline
\end{tabular}

$^{(a)}$ Number of sources. \\
$^{(b)}$ Actual mean distance to nearest neighbour of the sample and its 
          standard deviation. Since the sources in these samples are 
          distributed rather irregularly, the standard deviation occasionally 
          exceeds the mean. \\
$^{(c)}$ Active Galaxies are galaxies with unusually high X-ray or radio
          (few) luminosity. These do not overlap with any other subsample
          in this table (see text).
\vspace{-0.2cm}
\end{table*}

The  ``Catalogue of Quasars and Active Nuclei'' (V\'eron-Cetty \& V\'eron 
1996) is divided in three parts,--- quasars, BL Lac objects and AGN.
In addition to these three subsamples we shall consider
certain subsamples, e.g.\ sources selected by spectral type.
Because synchrotron emission is a favoured mechanism to explain
GRBs, we used criteria which are thought to favour the presence 
of jets. The resulting samples are listed in Table 2 together with
the sample size and the mean distance to the nearest neighbour. 
The selection criteria for the subsamples chosen are described 
in the following.

\subsubsection{Quasars}

We adopt the definition by V\'eron-Cetty \& V\'eron (1996) of a
quasar as a star-like object or an object with a star-like nucleus,
brighter than absolute visual magnitude $-23$. In addition to the
total quasar sample we consider three subsamples. The first is given
by objects characterized by a high optical polarization,
$P > 3$\%, which is mostly interpreted as evidence
for synchrotron radiation connected with a Doppler-boosted nuclear
jet (e.g.\ Wills et al.\ 1992).
The other two classes are defined by their radio-loudness.
Although the radio energy band is far away from the gamma-ray band 
discussed here, these subsamples are justified because of the 
strong correlation found between the source properties in
the radio and X-ray region. Already Einstein data showed a correlation 
between radio-loudness and both X-ray luminosity and X-ray spectral index 
(Wilkes \& Elvis 1987). This result was confirmed by ROSAT measurements
based on larger samples (see e.g. Schartel et al.\ 1996). 
In particular, the different spectral
properties may indicate that the X-ray spectra are
dominated by different physical mechanisms which may result
in a different behavior at higher energies.
We define the radio-loudness based on the radio-to-optical spectral index
(Stocke et al. 1985; Kellermann et al.\ 1989)

\begin{eqnarray}
\alpha_{RO} & = &  \frac{1}{5.38} \;
                   \log \left( \frac{S_{\rm 5 GHz}}{S_{2500\AA}}\right)
                   \nonumber \\
            & = &  \frac{1}{5.38} \; \left[ \; \log S_{\rm 5 GHz} - 23.0
                   + 19.76 + 0.4 V \right. \nonumber \\
            &   &  \;\;\;\;\;\;\;\;\;\; \left.
                    - 0.3 \log \left( z+1 \right) \right]
                    \nonumber
\end{eqnarray}
where the spectral fluxes are given in units of Jy and V is the object's
apparent  visual  magnitude.
Note that this definition is independent of optical luminosity. 

The radio-to-optical index has been K-corrected
assuming a spectral index of $\alpha_{R}=-0.7$ in the radio range
and $\alpha_{O}=-1.0$ in the optical regime.
The radio fluxes and the visible magnitudes were
taken from V\'eron-Cetty \& V\'eron (1996).
We define radio-loud quasars as those with $\alpha_{RO} > 0.3$.
A source is defined to be radio-quiet if either $\alpha_{RO} < 0.3$ or
no radio flux is given in the catalogue.

\subsubsection{BL Lac objects}

The V\'eron-Cetty \& V\'eron (1996) compilation provides
225 confirmed, probable or possible BL Lac objects.
In addition to the total sample and the
sample restricted to the optically {\it confirmed} BL Lacs
we consider three further subsamples.
The first one contains all known BL Lac objects which show an
optical continuum polarization $P > $3\% and the second one 
(called ``radio-selected'') is given by all objects with a 
flux density $>$ 0.1\,Jy at either $\lambda$ = 6 or 11\,cm.
Finally, we select all sources not fulfilling
one of the above-mentioned criteria. This latter sample is dominated by
BL Lac objects which are strong X-ray emitters.  
For 66 of the 82 objects in this group the mere designation 
already indicates that they were
discovered during X-ray missions.

\subsubsection{AGN}

The V\'eron-Cetty \& V\'eron (1996) compilation lists 2833 AGN.
By definition of the authors, all of them are fainter than absolute
magnitude --23.
This sample includes sources with spectra classified as
Seyfert 1, Seyfert 2  or  LINERs following the definition of Heckman (1980).
Further 97  sources are classified as probable or possible Seyfert galaxies.

The list of AGN includes 116 galaxies with a nuclear HII region.
These are now interpreted as a burst of star formation, but
were called Seyferts in the past and later were reclassified.
Therefore, we treat them separate from the Seyferts here.
We collect all objects without any morphological type assigned 
to them by V\'eron-Cetty and V\'eron (1996) into a class we term
``Active Galaxies'' (hereinafter AG). These typically consist of
galaxies with nuclear radio or X-ray sources. 
Note that we use the term ``AGN'' only for the combined
sample of Seyfert galaxies and LINERs. 
We also divide the AGN into radio-loud and radio-quiet samples,
defined by their radio-to-optical index $\alpha_{OR}$, in the same 
way as above for the quasars.  

\section{Search for Coincidences}

\subsection{Positional Coincidence Analysis}

In order to test our results if only for internal consistency, 
we compare the subsamples of QSOs and AGN defined above with three
samples of GRBs containing bursts up to a given positional error radius
(see Table 1).
The results are listed in Table 2. For every comparison we give the
expected chance coincidences $<C_B>$ derived from 1000 
randomly simulated GRB samples. 
Further we characterize the distribution of $C$
by its standard deviation. Since the distribution of coincidences is
not Gaussian, we also give the number of coincidences obtained with
the real GRB samples, $C_{1\fdg7}$, $C_{1\fdg8}$, and $C_{2\fdg0}$,
together with the probability P to measure a lower value than the
one obtained. Histograms of the distribution of $C$, i.e.\ the number of
coincidences for the artificial sets of GRBs (null hypothesis), 
are shown in Fig. 3 for all analyzed subclasses.
The expected value of coincidences (or the mean of the distribution), $<C_B>$, 
is shown as dotted line and the obtained value of real coincidences, 
$C_{1\fdg7}$ is indicated by a dashed line.

Three classes, the ``active galaxies'', all quasars and the radio-quiet
quasars show coincidences in excess of the expected value at the
90\%-level or more.
Since the radio-loud quasars alone do not show any excess coincidences,
we conclude that only radio-quiet quasars may be related to GRBs.
All other classes are compatible with the null hypothesis of
no relation with GRBs. In particular, the highly-polarized and the
radio-loud subsamples show no significant probability of being
associated with GRBs.

\subsection{Radio-quiet Quasars and Active Galaxies}

Only two classes, active galaxies and radio-quiet quasars show
coincidences in excess of random expectation above the 95\%-level.
The sample sizes range from 71 to 7146 members and the mean distance
to the nearest neighbour within a given sample varies from 0\fdg4
to $11\degr$ (see Table2). In this section we will test
whether any excess correlation may be influenced by the size of
the comparison sample and/or the distance to the nearest neighbour.
The expected number of chance coincidences was obtained by comparing
the positions of the sources with randomly simulated positions of the
GRBs (as explained in Sect. 2). Therefore, our results
are robust against large-scale structure in the burst locations.
A possible influence of any small-scale structure in the burst locations
remains to be checked.
We will also study the effect of the non-homogeneous distribution of objects
in the comparison samples.
In summary, we have to check for any hidden, yet unreported
characteristics in the sky distribution of GRBs and/or
in the sky distribution of AGN and QSOs which may be responsible
for the high levels of correlation with GRBs.

\begin{figure*}[tp]
\psfig{figure=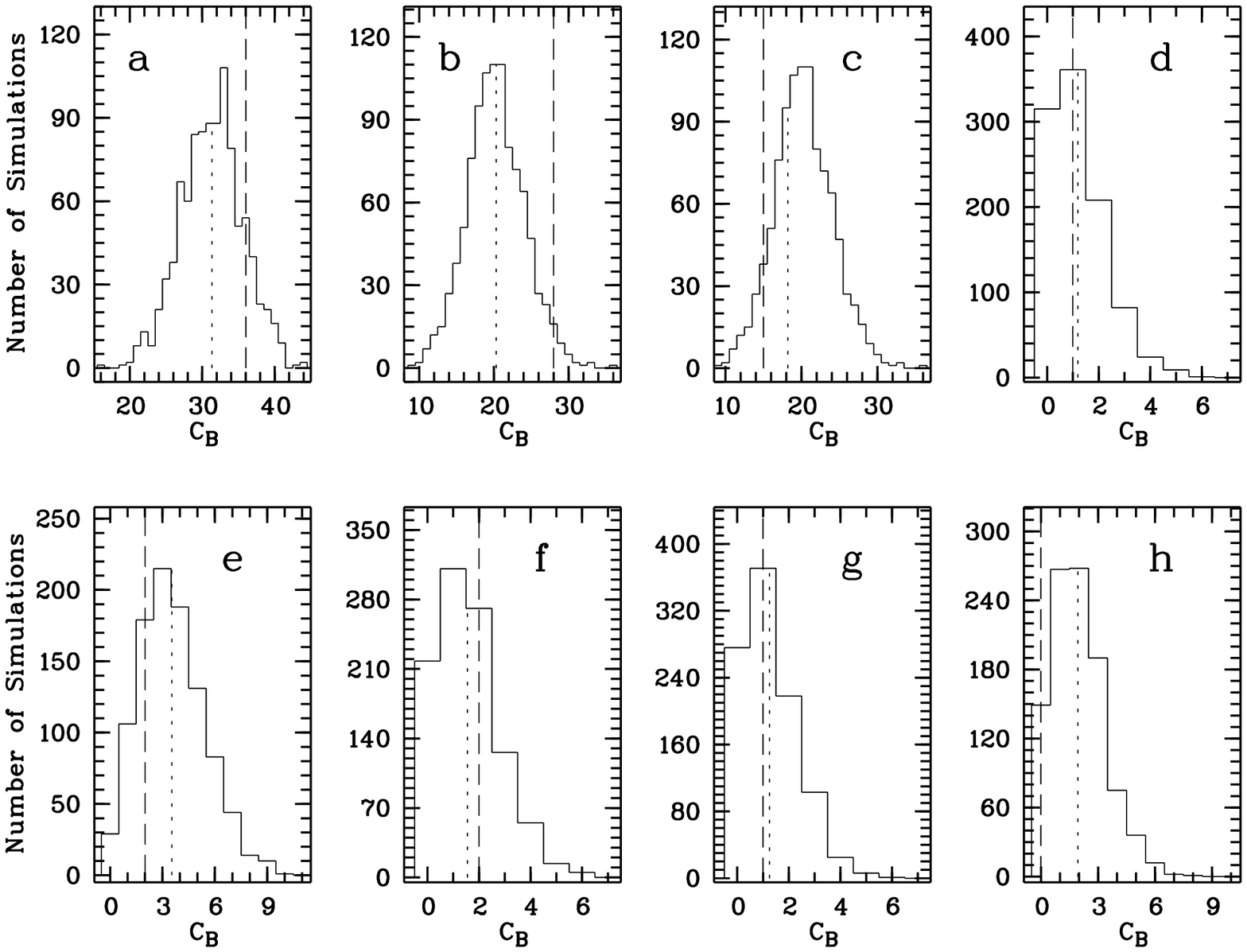,width=17.7cm
       ,bbllx=1.0cm,bblly=0.0cm,bburx=27.0cm,bbury=16.2cm,clip=}
\psfig{figure=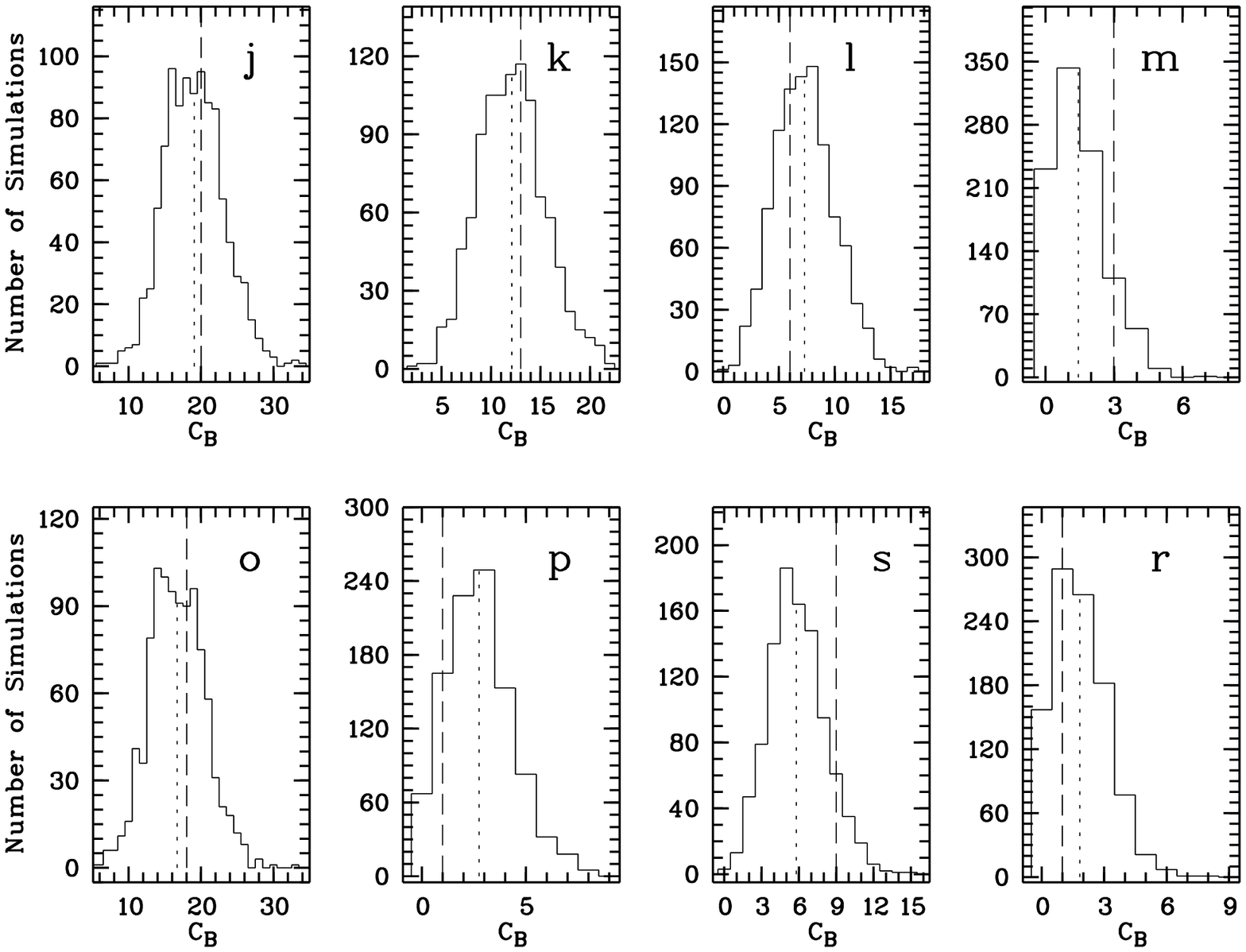,width=17.7cm
       ,bbllx=1.0cm,bblly=0.0cm,bburx=27.0cm,bbury=16.2cm,clip=}
      \vspace{-0.75cm}
      \caption{Distribution of the position coincidences for
               1000 random simulations of the 80 burst positions
               with $R<$1\fdg7. The dotted line shows the expected
               value for the null-hypothesis and the broken line
               the number of real coincidences between the
               gamma-ray burst positions and positions of quasars
               and AGN defined by different samples: The
               distribution is given for: quasars in the
               first panel: (a) all, (b) radio-quiet,
               (c) radio-loud, and (d) highly polarized quasars,
               for BL Lac objects in the second panel:
               (e) all, (f) confirmed, (g) highly polarized,
               (h) radio-selected, and (i) X-ray selected,
               for AGN in the third and fourth panel:
                (j) all, (k) Seyfert 1,
                (l) Seyfert 2, (m) probable Seyferts and (n)
                Liners, (o) radio-quiet, (p) radio-loud. Finally, the
                distribution is shown for active galaxies (s) and
                Nuclear H II regions (r). A detailed description of
                the selection criteria is given in the text.}
\end{figure*}

\begin{table*}[tp]
  \caption{Positional coincidences of redshift subsamples with randomly 
simulated GRB samples.}
\begin{tabular}{crrcrrrr} \hline \hline 
\noalign{\smallskip}
\multicolumn{2}{c}{Redshift} & \multicolumn{1}{c}{Number}
     & \multicolumn{1}{c}{Distance to} & \multicolumn{2}{c}{$R_t$=1\fdg7} &
       \multicolumn{2}{c}{$R_t$=1\fdg8} \\
& & \multicolumn{1}{c}{of} & \multicolumn{1}{c}{next Source} &
        $<C_B>$ & \multicolumn{1}{c}{$C_{1\fdg7}/P(\%)$} &
        $<C_B>$ & \multicolumn{1}{c}{$C_{1\fdg8}/P(\%)$} \\
\multicolumn{1}{c}{Range} 
& \multicolumn{1}{c}{$\bar{z}$}
& \multicolumn{1}{c}{sources}
& \multicolumn{1}{c}{$\bar{D}$ ($\degr$)} \\ 
\noalign{\smallskip}
\hline 
\noalign{\smallskip}
\multicolumn{8}{c}{Radio-quiet Quasars} \\
 0.00---0.58 &  0.35 &$883 $& $ 2.1\pm2.1 $ &
                $  9.7 \pm 3.0 $&$ 17 $ / $ 98.1  $&
                $ 16.8 \pm 3.9 $&$ 28 $ / $ 99.5  $ \\
 0.58---1.03 &  0.81 &$889 $& $ 1.3\pm1.7 $ &
                $  6.2 \pm 2.4 $&$ 13 $ / $ 99.1  $&
                $ 10.8 \pm 3.2 $&$ 22 $ / $ 99.8  $ \\
 1.03---1.41 &  1.24 &$903 $& $ 1.1\pm1.9 $ &
                $  5.5 \pm 2.3 $&$  9 $ / $ 90.0  $&
                $  9.6 \pm 3.1 $&$ 12 $ / $ 73.4  $ \\
 1.41---1.73 &  1.57 &$895 $& $ 1.1\pm1.8 $ &
                $  5.4 \pm 2.2 $&$  8 $ / $ 82.7  $&
                $  9.3 \pm 2.9 $&$ 11 $ / $ 66.4  $ \\
 1.73---2.00 &  1.89 &$871 $& $ 1.2\pm2.2 $ &
                $  5.8 \pm 2.2 $&$  7 $ / $ 66.8  $&
                $  9.9 \pm 2.9 $&$ 12 $ / $ 73.0  $ \\
 2.00---2.21 &  2.11 &$911 $& $ 1.3\pm2.2 $ &
                $  6.0 \pm 2.3 $&$  8 $ / $ 75.9  $&
                $ 10.5 \pm 3.0 $&$ 11 $ / $ 50.9  $ \\
 2.21---2.50 &  2.33 &$887 $& $ 1.2\pm1.8 $ &
                $  6.1 \pm 2.3 $&$  6 $ / $ 41.8  $&
                $ 10.5 \pm 3.1 $&$ 12 $ / $ 63.0  $ \\
 z $>$ 2.50  &  2.98 &$881 $& $ 1.4\pm2.0 $ 
                &  $  6.8 \pm 2.6 $&$  8 $ / $ 63.2  $&
                $ 11.9 \pm 3.3 $&$ 16 $ / $ 85.5  $ \\
\multicolumn{8}{c}{Active Galaxies} \\ 
z $<$   0.17  &  0.07    &$590  $&  $ 1.4\pm3.4 $ &
                $  3.5 \pm 1.8 $&$  6 $ / $ 87.6  $&
                $  6.1 \pm 2.4 $&$ 12 $ / $ 97.4  $ \\
z $\ge$  0.17 &  0.40    &$575  $&  $ 1.5\pm2.7 $ & 
                $  4.2 \pm 2.0 $&$  7 $ / $ 87.0  $&
                $  7.2 \pm 2.6 $&$ 14 $ / $ 98.9  $ \\
\noalign{\smallskip} 
\hline
\end{tabular}
\end{table*}

As mentioned above the position errors of GRBs are a function of the
measured $\gamma$-ray flux above 25 keV.
A restriction to bursts with the smallest errors leaves us
with the brightest bursts and thus  more coincidences
for nearby GRB sources are expected.
Therefore, sample selection by redshift offers an additional  possibility
to test our result, as we would expect more coincidences for the low-redshift
sources.
The main advantage of redshift-selected subsamples is given
through their intrinsic parameters (for example the mean distance to the
next source) which are different from the total sample.
Therefore, they allow to test our results against these parameters.
Table 3 lists the redshift bins, their mean redshift,
the number of sources in each bin,
and the mean distance to the nearest neighbour.

\subsubsection{Distance to the nearest neighbours}

Table 2 shows that comparison samples with the highest correlation with GRBs
(with the exception of the total quasar sample) have the smallest 
mean distance to the nearest neighbour. Active galaxies have a mean distance 
of 1\fdg0$\pm$2\fdg1 to their nearest neighbour and
radio-quiet quasars have only 0\fdg4$\pm$0\fdg9.
The subsamples defined by redshift allow to increase significantly the
distance to the nearest neighbour, as is obvious from Table 3.
In particular, in the lowest redshift sample of  radio-quiet
quasars a mean distance of 2\fdg1$\pm$2\fdg1 is achieved.
Samples with comparable values of nearest neighbour distance, e.g.\ all AGN 
(2\fdg1$\pm$1\fdg7), radio-quiet AGN (2\fdg2$\pm$1\fdg8) and
radio-loud quasars (2\fdg3$\pm$1\fdg6) do not show coincidences with GRBs
in excess of random expectation. We correlated the redshift-selected samples 
of QSO/AGN with randomly simulated versions (see Sect.\,2) of the two GRB 
samples with $R_t <$ 1\fdg7 and $R_t <$ 1\fdg8.
The results are given in Table 3.
For the two subsamples of radio-quiet quasars with lowest redshift,
probabilities of $> 98$\% are found for a correlation with both GRB samples.
Such a positive result is not obtained for
the ``active galaxies'', which show probabilities as low as 87\% for the 
GRB sample with the smallest error radii.

The $z \le 0.58$ subsample for the radio-quiet quasars shows
that our result is robust against an increase of the mean
distance to the nearest neighbour. On the other hand, the higher-redshift
subsamples demonstrate that even samples with a small distance to the nearest
neighbour may well yield negative results.

\subsubsection{Small-scale structure in the GRB distribution}

In order to test the positive result obtained for radio-quiet quasars and 
to investigate in more detail the results obtained for active galaxies
for a possible bias by the small-scale structure of the GRB
distribution, we tested our results using {\it shifted GRB samples}
(cf.\ Sect.\,2.3).
This covers two different effects: First, local differences in the BATSE
exposure time of certain sky regions causing fluctuations of the local 
GRB density are smeared out. Second, any possible internal structure of 
the GRB distribution (e.g. similar to the Great Wall structure of galaxies)
is tested.

The results are presented in the usual way in Table 4.
They strengthen the results obtained in the previous section.
In the first two redshift subsamples ($z<0.58$ and $0.58 < z < 1.03$)
the radio-quiet quasars reach probability levels of $>$99\% for correlation
with GRBs for both burst samples.  In contrast, the two subsamples of the
active galaxies do not exceed  a probability of 94\% for correlation with
GRBs.

\begin{table}[thpb]
\vspace{-0.2cm}
  \caption{Redshift Subsamples / Shifted comparison samples}
\begin{tabular}{crrrr} \hline \hline 
\noalign{\smallskip}  
\multicolumn{1}{c}{Redshift} &
\multicolumn{2}{c}{$R_t$=1\fdg7} &
\multicolumn{2}{c}{$R_t$=1\fdg8} \\
\multicolumn{1}{c}{Range} &
$<C_B>$ & $\!\!\!C_{1\fdg7}$/P(\%) &
$<C_B>$ & $\!\!\!C_{1\fdg8}$/P(\%) \\ \noalign{\smallskip} \hline 
\noalign{\smallskip}
\multicolumn{5}{c}{Radio-quiet Quasars} \\
 0.00---0.58 & $  9.6 \pm2.8  $&$ 17 $ / $ 99.8  $&
               $ 16.6 \pm3.9  $&$ 28 $ / $ 99.7  $\\
 0.58---1.03 & $  6.0 \pm2.0  $&$ 13 $ / $ 99.9  $&
               $ 11.0 \pm2.9  $&$ 22 $ / $ 100.  $\\
 1.03---1.41 & $  5.2 \pm2.0  $&$  9 $ / $ 94.2  $&
               $  9.6 \pm2.9  $&$ 12 $ / $ 74.9  $\\
 1.41---1.73 & $  4.8 \pm2.1  $&$  8 $ / $ 89.6  $&
               $  8.9 \pm3.0  $&$ 11 $ / $ 69.3  $\\
 1.73---2.00 & $  5.2 \pm2.1  $&$  7 $ / $ 70.9  $&
               $  9.5 \pm3.1  $&$ 12 $ / $ 74.7  $\\
 2.00---2.21 & $  5.5 \pm2.2  $&$  8 $ / $ 81.2  $&
               $  9.9 \pm3.3  $&$ 11 $ / $ 59.7  $\\
 2.21---2.50 & $  5.2 \pm2.3  $&$  6 $ / $ 57.7  $&
               $ 10.1 \pm3.2  $&$ 12 $ / $ 67.9  $\\
z $>$ 2.50    & $  6.1 \pm2.5  $&$  8 $ / $ 73.6  $&
               $ 11.6 \pm3.1  $&$ 16 $ / $ 87.9  $\\
\multicolumn{5}{c}{Active Galaxies} \\
z $<$   0.17 & $  3.6\pm1.9   $&$   6$ / $ 83.8  $&
               $  7.6\pm2.3   $&$  12$ / $ 93.4  $ \\
z $\ge$ 0.17 & $  4.2\pm 2.0  $&$   7$ / $ 88.8  $&
               $  8.4\pm 2.7  $&$  14$ / $ 93.7  $ \\
\noalign{\smallskip}
\hline 
\end{tabular}
\vspace{-0.2cm}
\end{table}

\begin{table}[tbhp]
  \vspace{-0.2cm}
  \caption{Redshift Subsamples /  Singular Coincidences} 
\begin{tabular}{crrrr} 
\hline \hline 
\noalign{\smallskip}  
\multicolumn{1}{c}{Redshift} &
\multicolumn{2}{c}{$R_t$=1\fdg7} &
\multicolumn{2}{c}{$R_t$=1\fdg8} \\
\multicolumn{1}{c}{Range} &
$<C_B>$ & \multicolumn{1}{c}{$\!\!\!C_{1\fdg7}$/P(\%)} &
$<C_B>$ & \multicolumn{1}{c}{$\!\!\!C_{1\fdg8}$/P(\%)} \\ 
\noalign{\smallskip} \hline 
\noalign{\smallskip}
\multicolumn{5}{c}{Radio-quiet Quasars} \\
 0.00---0.58 & $  6.9 \pm 2.6  $&$ 14 $ / $ 99.0  $&
               $ 11.8 \pm 3.4  $&$ 23 $ / $ 99.6  $ \\
 0.58---1.03 & $  3.7 \pm 1.9  $&$  8 $ / $ 95.9  $&
               $  6.4 \pm 2.5  $&$ 15 $ / $ 100.  $ \\
 1.03---1.41 & $  2.9 \pm 1.7  $&$  3 $ / $ 43.8  $&
               $  5.1 \pm 2.2  $&$  4 $ / $ 24.9  $ \\
 1.41---1.73 & $  2.9 \pm 1.7  $&$  1 $ / $  6.0  $&
               $  5.0 \pm 2.2  $&$  3 $ / $ 11.1  $ \\
 1.73---2.00 & $  3.1 \pm 1.7  $&$  4 $ / $ 63.2  $&
               $  5.4 \pm 2.3  $&$  7 $ / $ 72.4  $ \\
 2.00---2.21 & $  3.3 \pm 1.7  $&$  3 $ / $ 32.8  $&
               $  5.9 \pm 2.3  $&$  5 $ / $ 28.4  $ \\
 2.21---2.50 & $  3.5 \pm 1.7  $&$  2 $ / $ 12.4  $&
               $  6.0 \pm 2.4  $&$  5 $ / $ 27.0  $ \\
 $>$ 2.50    & $  4.0 \pm 2.0  $&$  5 $ / $ 62.8  $&
               $  7.0 \pm 2.6  $&$  8 $ / $ 61.1  $ \\
\multicolumn{5}{c}{Active Galaxies} \\
z $<$   0.17 & $  1.7 \pm 1.3  $&$  2 $ / $ 46.8  $&
               $  3.0 \pm 1.7  $&$  4 $ / $ 65.8  $ \\
z $\ge$ 0.17 & $  2.5 \pm 1.5  $&$  3 $ / $ 54.2  $&
               $  4.3 \pm 2.0  $&$  6 $ / $ 74.7  $ \\
\noalign{\smallskip}  
\hline 
\end{tabular}
  \vspace{-0.35cm}
\end{table}

\subsubsection{Non-Homogeneity of the Comparison Samples}

In Sect. 3.2.1 we were able to rule out a bias caused by the mean 
distance to the nearest neighbours for the radio-quiet quasars whereas 
our results for the active galaxies were not convincing.
We are aware that the non-homogeneity of the comparison sample is not 
sufficiently characterized by the mean distance to the nearest neighbours.
To test for a possible bias of this effect, we count only bursts with
only a single QSO or AGN in its 1\,$\sigma$ error circle 
(see { \it singular coincidences} in Sect. 2.4).
The expected chance coincidences and their distribution 
were obtained again with simulated GRB positions.

Table 5 gives the results and confirms a high probability for correlation
with GRBs in excess of random expectation for the two lowest redshift 
quasar samples, reaching a probability above the 99\% level.
In contrast to this, we do not find any support for a correlation with
GRBs for the ``active galaxies''.

\section{Results}

We first summarize all classes  for which no coincidences in excess
of random expectation were found. In the second subsection we will strengthen 
the case for a correlation between GRBs and radio-quiet quasars.

\begin{table}[tbhp] 
\vspace{-.2cm}
\caption{Upper Limits}
\begin{tabular}{llcrcrr} \hline \hline 
\noalign{\smallskip}
\multicolumn{1}{c}{Main}
&  \multicolumn{1}{c}{Subclass} & \multicolumn{3}{c}{Number}
& \multicolumn{1}{c}{$U_{3\sigma}$}
& \multicolumn{1}{c}{$R_{U_{3\sigma}}^{(a)}$}\\
\multicolumn{1}{c}{Class} &
\multicolumn{1}{c}{or}           & \multicolumn{3}{c}{of} & \\
& \multicolumn{1}{c}{selection}    & \multicolumn{3}{c}{sources}   & \\ 
\noalign{\smallskip}
\hline 
\noalign{\smallskip}
QSO      & radio-loud         &&$1377$&&$ 18.2  $&$   1.50 $\\
QSO     & highly polarized    &&$  72$&&$  5.2  $&$   8.17 $\\
BL      & all                 &&$ 220$&&$  4.3  $&$   2.21 $\\
BL      & confirmed           &&$  93$&&$  4.0  $&$   4.87 $\\
BL      & highly polarized    &&$  76$&&$  4.3  $&$   6.40 $\\
BL      & radio-selected      &&$ 119$&&$  3.3  $&$   3.14 $\\
BL      & X-ray selected      &&$  82$&&$  4.1  $&$   5.66 $\\
AGN     & AGN                 &&$1553$&&$ 16.8  $&$   1.22 $\\
AGN     & Seyfert 1           &&$ 888$&&$ 13.1  $&$   1.67 $\\
AGN     & Seyfert 2           &&$ 496$&&$  9.8  $&$   2.24 $\\
AGN     & Probably Seyfert    &&$  97$&&$  5.2  $&$   6.07 $\\
AGN     & LINER               &&$  71$&&$  3.1  $&$   4.94 $\\
AGN     & radio-quiet         &&$1346$&&$ 16.9  $&$   1.42 $\\
AGN     & radio-loud          &&$ 166$&&$  2.5  $&$   1.70 $\\
        & Active galaxies (AG)&&$1165$&&$ 16.9  $&$   1.64 $\\
        & Nuclear H II region &&$ 116$&&$  2.9  $&$   2.83 $\\
\noalign{\smallskip}
\hline
\label{ul}
\end{tabular}
\vspace{-0.2cm}

$^{(a)}$: per source in 10$^{-2}$ yr$^{-1}$
\end{table}

\subsection{Upper Limits}

Quasars and AGN are often discussed in the frame of the
standard model assuming a massive compact object in the center,
most probably a super-massive black hole.
Many spectral features, like optical polarization or
radio-loudness, are interpreted as evidence of a nuclear jet.
Most of the samples and subsamples of QSO and AGN analyzed here
show no evidence for a correlation with GRBs. However,
the measured values can be used to estimate upper limits for the emission
of GRBs. To determine the 3\,$\sigma$ upper limit we use always our analysis +
the $R_t <$ 1\fdg8 burst sample which reaches the highest significance.
The 3\,$\sigma$ upper limit is given by:
\[
U_{3\sigma} = C_{1\fdg8} + 3 \Delta C_{1\fdg8} - <C_{1\fdg8}>
\]
\noindent
where $C_{1\fdg8}$ is the number of coincidences with the real GRBs, and
$<C_{1\fdg8}>$ is the number of coincidences expected for a simulated
burst sample. $ \Delta C_{1\fdg8}$ is approximated by the
standard deviation of $<C_{1\fdg8}>$ as provided in Table 2,
together with the other two values. The resulting upper limits
are listed in Table \ref{ul}.
As a quantity independent of the amount of members in the different
subclasses we define the emission rate, $R_{U_{3\sigma}}$:
\begin{eqnarray}
R_{U_{3\sigma}} & = & (1/0.38)~(1/0.68)~(1/3.42)~U_{3\sigma} N^{-1}\;\;per\;\;year 
\nonumber \\
                & = & 1.132~U_{3\sigma} N^{-1}\;\;per\;\;year\;\;\;. 
\end{eqnarray}

\begin{table*}[thpb]
  \vspace{-0.2cm}
  \caption{Magnitude subsamples of radio-quiet quasars}
\begin{tabular}{cccccrrrr} \hline \hline \\
\multicolumn{2}{c}{Magnitude} &
\multicolumn{3}{c}{Number}    &
\multicolumn{2}{c}{$R_t$=1\fdg7} &
\multicolumn{2}{c}{$R_t$=1\fdg8} \\
\multicolumn{2}{c}{ } &
\multicolumn{3}{c}{of} \\
\multicolumn{1}{c}{Range} &
\multicolumn{1}{c}{$\bar{M_{abs}}$} &
\multicolumn{3}{c}{sources} &
$<C_B>$ & \multicolumn{1}{c}{$C_{1\fdg7}/P(\%)$} &
$<C_B>$ & \multicolumn{1}{c}{$C_{1\fdg8}/P(\%)$} \\ \\ \hline \\
--23.0 --- --24.1   &$ -23.6\pm0.3 $&& 836 &&
                   $  7.3\pm2.6 $&$  9 $ / $ 67.4  $&
                   $ 12.6\pm3.3 $&$ 18 $ / $ 92.4  $ \\
--24.2 --- --25.0   &$ -24.6\pm0.3 $&& 847 &&
                   $  5.9\pm2.3 $&$ 13 $ / $ 99.4  $&
                   $ 10.2\pm3.0 $&$ 18 $ / $ 99.2  $ \\
--25.1 --- --25.6   &$ -25.4\pm0.2 $&& 784 &&
                   $  5.2\pm2.2 $&$ 9 $ / $ 91.9   $&
                   $  8.9\pm2.9 $&$ 12 $ / $ 81.9   $ \\
--25.7 --- --26.1   &$ -25.9\pm0.1 $&& 820 &&
                   $  4.8\pm2.2 $&$ 8 $ / $  88.0   $&
                   $  8.3\pm2.8 $&$ 15 $ / $ 98.2   $ \\
--26.2 --- --26.6   &$ -26.4\pm0.1 $&& 910 &&
                   $  5.7\pm2.2 $&$ 9 $ / $  90.2   $&
                   $  9.7\pm3.0 $&$ 17 $ / $ 97.9   $ \\
--26.7 --- --27.1   &$ -26.9\pm0.1 $&& 875 &&
                   $  6.1\pm2.3 $&$ 11 $ / $ 96.8   $&
                   $ 10.6\pm3.1 $&$ 16 $ / $ 93.3   $ \\
--27.2 --- --27.9   &$ -27.5\pm0.2 $&& 993 &&
                   $  7.3\pm2.6 $&$ 10 $ / $ 81.4   $&
                   $ 12.6\pm3.3 $&$ 15 $ / $ 71.9   $ \\
--28.0 --- --31.9   &$ -28.8\pm0.7 $&& 871 &&
                   $  8.9\pm3.0 $&$ 7 $ / $ 21.1   $&
                   $  15.6\pm3.7 $&$ 14 $ / $ 29.9   $ \\
 \\ \hline
\end{tabular}
\end{table*}

\noindent 
This rate is given by  the derived upper limits divided by 
the size $N$ of the comparison sample, and  by the observation time
of BATSE (3.42 years). 
Further it is corrected for the  68\% confidence interval we used, 
and the probability of detecting a burst (38\%, cf. \ Meegan et al. 1996).
All numbers refer to the 134 GRBs in the 3B catalogue
having $R_t <$ 1\fdg8.

Our initial positive result for a correlation of GRBs with ``active galaxies''
(AG) was not confirmed in the previous chapter. Thus we give an
upper limit for AGs in Table \ref{ul}. The positive result obtained initially
might be due to the non-homogeneity of the sky distribution of AG.
However, an additional bias due to a small-scale structure of the
GRB distribution can not be excluded at this point.

\subsection{Radio-quiet Quasars}

Because of the inverse relation between burst flux and positional
error radius we expected to strengthen 
the result obtained in our first search for correlation
for the low-redshift quasars.  In Sect. 3.2. we were able to show 
that the result is influenced neither by the distance to the nearest neighbour, 
nor the small-scale structure of the burst locations, nor by the
non-homogeneity of their sky distribution. In this
section we will try to strengthen our positive correlation
for radio-quiet quasars.

Seyfert galaxies are closely related to radio-quiet quasars. 
Because many radio-quiet quasars show a Seyfert-like spectrum,
it appears that the main difference between these two classes
is the absolute magnitude. 
Given the negative result obtained for the Seyfert galaxies,
it seems reasonable to presume that the absolute magnitude influences
the emission rate of GRBs.  
Therefore, we also defined subsamples according to absolute magnitude
as listed in the compilation of V\'eron-Cetty \& V\'eron (1996).
Again, the values obtained were compared with randomly simulated GRB
samples and the results are provided in Table 7.
The probability levels do not support coincidences in excess of
random expectation for the fainter magnitudes
($-23.0 > \bar{M}_{abs} > -24.1 $)
at more than the 95\%-level. However, excess coincidences are indeed
found for quasars brighter than $M_{abs} = -24.2$. It appears that only 
intrinsically bright quasars show a detectable emission rate of GRBs.
This interpretation is in agreement with the negative result
found for the Seyfert galaxies.

\begin{table}[thpb]
  \vspace{-0.2cm}
  \caption{Quasar sample selected by magnitude and redshift}
\begin{tabular}{crrrr} \hline \hline 
\noalign{\smallskip}
\multicolumn{1}{c}{Method$^{a}$ } &
\multicolumn{2}{c}{$R_t$=1\fdg7} &
\multicolumn{2}{c}{$R_t$=1\fdg8} \\
\multicolumn{1}{c}{} &
$<C_B>$ & \multicolumn{1}{c}{$\!\!\!C_{1\fdg7}/P(\%)$} &
$<C_B>$ & \multicolumn{1}{c}{$\!\!\!C_{1\fdg8}/P(\%)$} \\ 
\noalign{\smallskip}
\hline 
\noalign{\smallskip}
\multicolumn{5}{c}{Radio-quiet Quasars with :  } \\
\multicolumn{5}{c}{Magnitude $\le$ -24.2 and $z < 1.0$ } \\
1  &$  8.4\pm2.8 $&$ 20 $ / $   100.0 $&
            $ 14.5\pm3.7 $&$ 27 $ / $   99.7 $ \\
2  &$  7.9\pm2.3 $&$ 20 $ / $  100.0 $&
            $ 14.0\pm3.5 $&$ 27 $ / $   100.0 $ \\
3  &$  5.4\pm2.3 $&$ 17 $ / $   100.0 $&
            $  9.2\pm3.1 $&$ 20 $ / $   99.8 $ \\
\noalign{\smallskip} 
\hline
\noalign{\smallskip}
\end{tabular}

$^{a}$ Numbers denote the different samples: 1 = simulated control sample; 
      2 = shifted control sample; 3 = singular coincidences    
\end{table}

We conclude that there is an increased probability for the {\it nearby} and
for the absolutely {\it brighter} radio-quiet quasars to correlate with GRBs.
It is therefore tempting to construct an ``optimized quasar sample'' with even
higher probability for emitting GRBs.
We defined such a sample by $z\le1.0$ and $M_{abs} \le -24.2$.
It contains 967 radio-quiet quasars and was tested with the
three methods described in Sect. 2, for the two GRBs samples with
the smaller error radii.   Results are provided in Table 8.
The corresponding distribution of the positional coincidences for the
null-hypothesis of the ``optimized quasar sample'' are plotted in Fig. 4.
The value expected at random is shown as a dotted line and the
value obtained for the real GRBs is indicated by a dashed line.
For all six comparisons the probability for a real correlation with GRBs
exceeds 99.7\%. Moreover, in four
of the six comparisons the 1000 simulations of GRBs positions did not show 
a single case with a number of coincidences as high as for the real GRBs.

\begin{figure*}[thbp]
\vspace{-0.2cm}
 \psfig{figure=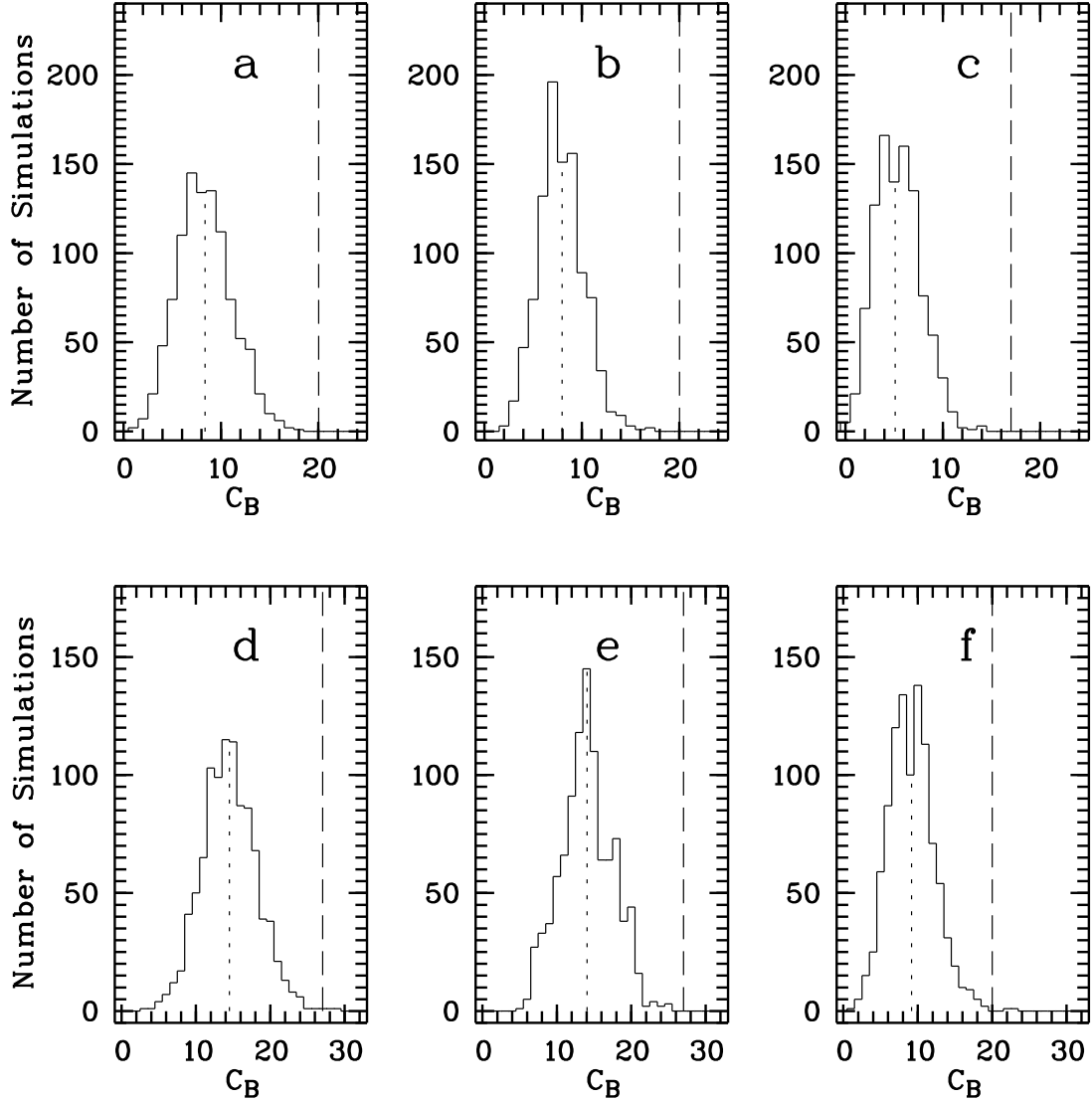,width=16.50cm
       ,bbllx=0.50cm,bblly=0.90cm,bburx=17.50cm,bbury=16.20cm,clip=}
      \vspace{-0.3cm}
      \caption{Distribution of the position coincidences of the
               null-hypothesis for the optimized quasar sample
               given by $z<1.0$ and $M_{abs} \le -24.2$. 
               The expected value of the background position 
               coincidences is indicated with a dotted line and the obtained 
               value of the real  coincidences is shown with a 
               dashed line. The upper  panel
               shows the comparison for 80 gamma-ray burst positions
               defined by R$<$1\fdg7 and the lower  for 132 burst positions
               defined by R$<$1\fdg8, respectively. In {\bf a} and {\bf d} the
               80 (132) positions were simulated randomly 1000 times.
               The distributions {\bf b} and {\bf e} were obtained by 
                shifting the
               original burst positions 1000 times. And finally, in
               {\bf c}  and {\bf f} only bursts are counted which show 
                not more than one coincidence.  }
\end{figure*}

Frequently, two catalogues of objects detected in different wavebands are 
cross-identified from a study of the radial histogram of the distances between
suspected counterparts. We did not use this method until this point
because many bursts have more than one QSO or AGN in their error circle, 
which would significantly increase the chance coincidences at 
large distances between GRBs and objects of the comparison sample.
However, the ``optimized quasar sample''
($z\le1.0$ and $M_{abs} \le -24.2$) allows to illustrate our result
with the radial histogram of the distances between
suspected counterparts.
The differential distribution of coincidences is given by the number of
quasars, $dN$ in an annulus between a radius $r$ and $r+dr$.
The distribution of the real coincidences is a function of the positional
errors which is not very well determined. However, the real coincidences 
must be located within a distance less than 3 times the positional error 
radius, $R_t$, assuming a two-dimensional Gaussian distribution.
The number of randomly expected coincidences is proportional to the
cumulative area of all annuli. Thus, the number of sources in a given 
annulus as a function of the radius is given by

\[ dN(r) = A(r, \Delta D) 2 \pi r dr + B 2 \pi r dr \]
\noindent where A(r) is the probability for real coincidences per unit area, 
and B is the probability for chance coincidences per unit area.
We divide by $r$ and set 

\[ C = A / ( 2 \pi) \;\;\;  {\rm and}  \;\;\; D = B / ( 2 \pi). \]

\noindent Thus, we may rewrite:

\[ \frac{dN(r)}{r} = C(r, \Delta D) dr   + D dr \]

\begin{figure}[thbp]
\vspace{-0.2cm}
\psfig{figure=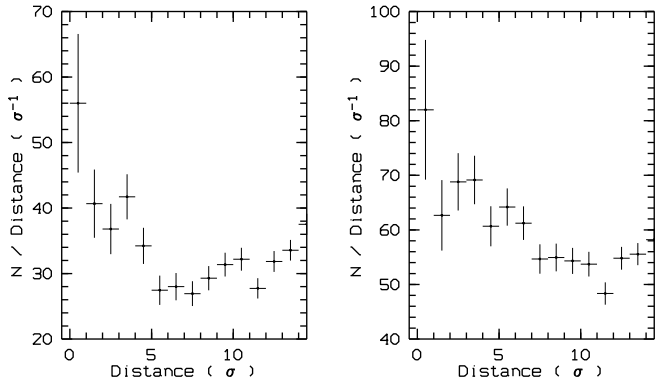,width=8.80cm
       ,bbllx=2.9cm,bblly=5.cm,bburx=19.0cm,bbury=14.4cm,clip=}
     \vspace{-0.2cm}
      \caption{The area-normalized  distribution of pairs of
            gamma-ray bursts and radio-quiet quasars
            as function of their mutual distance, scaled in units
            of the position uncertainty $\sigma$ of the bursts.
            The left panel shows only the N=80 GRBs with position uncertainty
            of $R_t <$ 1\fdg7. The right panel is based on 136 GRBs
            with $R_t <$ 1\fdg8. In both figures the sample of
            N = 967 radio-quiet quasars is defined by
            $z \le  1.0$ and $M_{abs} \le -24.2 $.  }
\end{figure}
  
To allow for the very different sizes of the GRB positional 
errors we scale the distance between GRB and radio-quiet quasars in units of
the positional error $\sigma$, i.e.\ we choose $dr=\sigma$  to obtain the
normalized differential distribution, which should be constant
for $r > 3 \sigma$. In Fig. 5 the
normalized differential distribution for the optimized quasar sample
is given.
In the normalized differential distribution every coincidence between a
gamma-ray burst and a quasar is counted.
This results in a doubling of the background value for
the first bin as opposed to counting  every GRB only once.
Nevertheless, the coincidences we find in excess of random expectation
between 2\,$\sigma$ and 5\,$\sigma$ more than compensate 
for the poor statistical significance of the first bin (cf.\ Fig.\,5).

By  substituting  $U_{3\sigma}$ with  ($C_{1\fdg8}-<C_{1\fdg8}>$)
in equation 2 we obtain  the emission rate for the detected 
radio-quiet quasars:
\[
R_{emission} = 1.132~(C_{1\fdg8}-<C_{1\fdg8}>) N^{-1}\;\;in\;\;y^{-1}
\]
Using the results obtained for the "optimized quasar sample" 
given in the first line of table 8 we determine an emission rate of burst of
0.0146$\pm$0.004 y$^{-1}$ 
per radio-quiet quasar.  

\section{Discussion}

\subsection{Correlation with radio-quiet quasars}

We have found evidence that intrinsically bright radio-quiet quasars
with $z<1.0$ coincide with GRBs from the 3B catalogue
at a confidence level of more than 99\%.

The correlation with low-redshift objects is not surprising.
For statistical reasons our study had to be restricted to 
GRBs  with the smallest positional errors, which tend to be
the brightest GRBs. These GRBs are 
preferably nearby since the luminosity distribution of GRBs
cannot be very flat (Hakkila \etal 1995),
and thus one introduces  a strong selection effect 
for low-redshift quasars. Based on the different methods which we applied 
to test our results, we are confident to have excluded any bias from an  
inhomogeneous sky distribution of quasars.
In addition, we showed our result to be robust against a possible
small-scale structure in the sky distribution of $\gamma$-ray bursts.

The restriction to intrinsically bright quasars with $M_{abs} < -24.2 $ 
seems to be a real physical selection indicating that the GRB
emission rate is at least lower for intrinsically faint quasars.
This interpretation is supported by the negative result obtained
for the Seyfert galaxies which are intrinsically fainter than
radio-quiet quasars but otherwise have very similar properties. 

We cannot exclude that other classes of AGN and QSOs emit bursts with a 
comparable rate since the derived 3\,$\sigma$ upper limits (see Table \ref{ul})
are above the emission rate of the radio-quiet quasars (0.0146 y$^{-1}$ per quasar). 
The lower emission rate of radio-loud quasars argues against a 
direct connection between radio jets and GRB emission.

\subsection{Comparison with previous results}

Two previous studies may appear in contradiction with our results.
However, a thorough analysis shows that one of these is not comparable with 
our study, and  that the other in fact does not contradict our results.

\subsubsection{The Kurt \& Komberg (1995) analysis}

Kurt \& Komberg (1995) compared the two-point angular correlation
function of 458 gamma-ray bursts with the two-point angular correlation
function of a sample of 413  unresolved extragalactic radio sources
(QSO, BL Lac and radio galaxies).
They found that the correlation function of GRBs does
not differ from a random distribution over all angular scales
while the one for the radio sources shows significant
autocorrelation on small angular scales at a level of $> 3\sigma$. 
Kurt \& Komberg conclude that GRBs do not belong to a normal population 
of galaxies with $z< 1.5$. This may not be correct in this generalization,
particularly since these authors dealt with compact radio sources and did 
not consider radio-quiet objects.
Moreover, Drinkwater \& Schmidt (1996) have recently
pointed out the very different clustering properties of radio
galaxies and radio-loud quasars, both of which are well-mixed in
 Kurt \& Komberg's sample.
Because we found only coincidences in excess of random expectation
for radio-quiet quasars, our result is not in contradiction to 
that of Kurt \& Komberg (1995). Furthermore, 
we found no hint for a correlation between radio-loud objects
and GRBs which agrees with  their conclusion.

\subsubsection{The Webber et al. (1995) correlation}

Webber et al. (1995) compared $\sim$60 GRBs located to within very
small error boxes of $\sim$0.25 deg$^2$ or less  with several catalogues 
of extragalactic objects.
More than half of their sample of GRBs was taken from the 
first Interplanetary Network (IPN) GRB Catalog (Atteia et al.\ 1987), 
seven GRBs from the COMPTEL catalogue of bursts (Hanlon et al.\ 1994) 
and 16 bursts from the third IPN (Hurley, private comm.).
They find that any correlation between these catalogued extragalactic  
objects and the GRBs are consistent with chance expectation.
In order to compare the bursts with the 
extragalactic objects they count the number of the closest 
bursts falling in bins of width 0\fdg5 in radial separation
independent of the actual shape of the error box.
They found two coincidences: QSO 2219$-$420 (z\,=\,1.3) in the error box
of GRB 790419 and QSO 0116--288 ($z=0.798$; which was discovered in a deep 
optical follow-up search of the GRB error box by Pedersen \etal 1983) in the 
error box of GRB 781119  (obviously mis-printed as GRB 791119 
in Webber et al. 1995). 
Webber et al. (1995) simulated 100 sets of quasar  positions and obtained a 
background value of  $3.7\pm1.8$.
Motivated by the large increase of known quasars since the 1989 compilation
of AGN and QSOs by V\'eron-Cetty \& V\'eron as used by Webber et al. (1995), we 
reanalyzed Webber's sample of GRBs with the 1995 compilation of radio-quiet QSO.
Because the error boxes of the seven bursts localized by COMPTEL
have areas well within $\sim$0.25 deg$^2$, which is larger than
most of the error boxes of the IPN GRB Catalog, we excluded the former here.
For the 28 bursts from the first IPN GRB Catalog (Atteia et al.\ 1987),
we compared the positions of radio-quiet QSO  
with the exact (elongated) shape of the error boxes and arrived at the same 
result as Webber et al. (1995): one quasar each is located in the 
error boxes of GRB 790419 and GRB 781119. Then we determined the expected 
background value with 1000 simulations for the 28 burst positions.
Accepting quasars within a bin width of 0\fdg5 we
derive the same result  as reported by Webber et al. (1995), namely that a 
correlation between quasars and the GRBs is consistent with chance expectation.
In a second approach, rather than using a standard radial distance of 0\fdg5
for all bursts, we converted the area of a given error box to a 
corresponding error circle for each burst. With this approach
we expect $0.18\pm0.4$  random coincidences as compared to the above value of 
$3.7\pm1.8$. Thus, the probability to
find two radio-quiet quasars in the error boxes of the GRBs is only 1.2\%.
This result corresponds to a probability of 98.8\% for a positive correlation,
i.e.\ much  higher than derived by Webber et al.\ (1995) and 
confirming our results. 

\subsection{Host questions}

Our result of a correlation between GRBs and intrinsically bright, 
radio-quiet quasars does not necessarily imply that GRBs are 
produced in the nuclei of these quasars, but possibly in their host
galaxies. Thus, there are two different questions
related to the hosts of GRB sources: Are host galaxies bright enough to be
detectable in previous or future deep searches? What do we know about the
host galaxies of radio-quiet quasars?

\subsubsection{Deep optical searches and the GRB hosts}

Earlier deep optical searches of small GRB error boxes (by imaging in 
bandpasses from U to K) did not find suitable host galaxies (Schaefer 1992).
This  has become known as the ``no-host" problem  for cosmological 
GRB scenarios (Fenimore \etal 1993a).
Recently, near-infrared imaging of six  small GRB error boxes 
(a sample of GRBs different from that used by 
Schaefer (1992)) in the JHK$'$ bands has shown that
the ``no-host" problem does not exist (Larson \etal 1996).
On the contrary, even an 
overabundance of brighter (K $\leq$ 15.5 mag) galaxies was found. 
Whether these might be hosts for radio-quiet quasars remains to be seen.

As the detectability of quasars is concerned, 
the lack of  excess quasars in GRB error boxes is 
neither contradicting our results nor surprising.
In photometric studies, radio-quiet quasars are not separated from 
other types of AGN or radio-loud quasars, and thus the existence of any
excess of radio-quiet quasars is easily missed.
We should note here that a recent deep optical imaging (Vrba \etal 1995)
has revealed marginal evidence for a quasar excess at the rate of one per
high Galactic latitude GRB error box. Further studies able
to distinguish  
between the different types of AGN are certainly needed to clarify
this situation.

In addition, there are two probable kinds of quasars which are difficult
to detect in optical, photometric searches.
First, there are only a few quasars showing Seyfert 2 type spectra.
If the ratio of Sy1/Sy2 type spectrum quasars were equally large
as the ratio of Sy1/Sy2 galaxies,
then a large number of quasars will have been missed in optical studies.
Ultra-luminous infrared galaxies discovered by IRAS
(Sanders et al.\ 1988) might be  a second possibility to explain
bursts lacking optical counterparts.
These are suggested to be young quasars in a dust-enshrouded
initial phase.

\subsubsection{The hosts of radio-quiet AGN}

Until quite recently radio-quiet quasars were thought to be associated
mainly with spiral galaxies, and radio-loud ones with elliptical
galaxies (see Bahcall et al. 1995 for an overview of references to
previous, apparently contradicting results).
However, very recent deep images taken with the {\it Hubble Space
Telescope} suggest that quasars reside in a wide variety of environments,
not confirming the above preconception. If radio-quiet quasars
are indeed equally likely to be associated with both spirals and
ellipticals, then their clustering properties are difficult to
predict. 
Together with the reasonable assumption that GRBs are an
extremely rare phenomenon in the life of a galaxy, the lack of
autocorrelation among GRBs seems more natural.

On the other hand, the production of GRBs must be related to some
specific property of the quasar, otherwise one would expect also 
a correlation of GRBs with normal spirals, which is not observed.
In the next section it is shown that the detected number of GRB 
and the total number of quasars is in agreement with the 
emission rate determined in section 4.2. Because the number
of spirals exceeds the number of quasars by a factor of 
order $10^3$, our result argues strongly against
a relation between GRB and normal spirals.

\subsection{Luminosity function of quasars}

In a Friedmann universe with zero cosmological 
constant the luminosity distance d$_l$ is defined as

\begin{displaymath}
d_L = \frac{c}{H_0 q_0^2} \left\{ q_0 z + \left(  q_0 - 1 \right) 
\left[ \left( 1 + 2 q_0 z  \right)^{\frac{1}{2}} - 1 \right] \right\} ,
\end{displaymath}

\noindent
whith $H_0$ the Hubble constant, $q_0$ the deceleration parameter
and $c$ the velocity of light. The comoving
volume $V(z)$ can be approximated by:

\begin{displaymath}
V (z) \approx \frac{4 \pi}{3} \left( \frac{d_L}{z+1} \right)^3 \;\;\;.
\end{displaymath}

\noindent 
For $q_0 = 0.5 $  this correspond to:

\begin{eqnarray*}
V (z) & \approx & \frac{32 \pi}{3} \left(\frac{c}{H_0}\right)^3
\left(\frac{\left( z + 1 - \sqrt{\left(z + 1\right)} \right)}{z+1} \right)^3 \\
& \approx &
7.24 \;\;\;10^{12}\;Mpc^{3}\; 
\left(\frac{\left( z + 1 - \sqrt{\left(z + 1\right)} \right)}{z+1} \right)^3
\;\;\;, \\
\end{eqnarray*}

\noindent
assuming  $ H_0 = 50$ Mpc$^{-1}$ km s$^{-1}$.
From Figure 2 in  Hewett et al. (1993) the cumulative space 
density of quasars $\Phi(<M_{B_J}, Mpc^{-3})$ for the 
optimized quasar sample can be estimated. 
Assuming $M_{B_J} \approx M_v < -24.1 $ we get $\Phi = 8\;10^{-8}\;Mpc^{-3}$ 
for the redshift range $ 0.2<z< 0.5$  and  
$2.2\;10^{-7}\;Mpc^{-3}$ for the redshift range 
$ 0.5<z< 1.0$, respectively.
To derive the total number of quasars we further assumed 
that $\Phi_{0.2<z<0.5}$ holds for redshifts below $z=0.2$ and that
15\% of the quasars are radio-loud.
Under these assumptions we can estimate that in total 2.9 10$^4$ quasars
meet the criteria of the optimized quasar sample. The number of quasars
studied, 967, therefore corresponds to 3.4\% of the total sample.
This number has to be compared with the percentage of GRBs 
cross-identified with a radio-quiet QSO.
According to Table 8 we found counterparts to 12.5 GRBs for a total
sample of 134 bursts studied, or a detection rate of 9.3\%$\pm$2.8\%. 
Thus, the expected detection rate is within two sigma of 
the observed one, a reasonable  agreement given the
small number statistics.

One should also be aware that there are two biasing effects which
increase the rate of optical counterparts. The first is that
more luminous quasars at a similar redshift show a higher
probability to be detected in most kinds of surveys.
Our finding that more luminous quasars have a higher probability of
being a GRB counterpart suggest that the basic quasar sample
used is already biased towards a higher rate of counterparts.
A further bias towards a higher rate of counterparts is caused by
the clustering of radio-quiet quasars (Shanks \& Boyle 1994).
Because the clustering occurs on small scales it does not affect our
results given the large error boxes of the GRBs. On the other
hand, unknown quasars show a higher probability to
be located near to a known quasar in comparison to an arbitrary
location on the sky.

\subsection{Models for GRBs from quasars}

Merging neutron stars and black holes are one possibility to explain 
quasars as counterparts of GRBs and are  discussed in the first of the 
following subsections. There are not many publications explaining GRB with 
emission by the active nucleus itself.
Because our result strongly suggests such a mechanism we show  
in a second subsection  that the gamma-ray emission of radio-quiet 
quasars/AGN can reach the energy range of GRBs. 
Finally, we discuss the model of Leiter (1980) which may  
explain the observed correlation.

\subsubsection{Merging neutron stars and black holes}

An interesting scenario to explain GRB from AGN and quasars was proposed by 
Epstein et al. (1993). Because compact-star binaries have a higher mass
than normal stars they settle to the center of the nuclear stellar cluster
(fraction of a parsec) before they coalesce. 
The binary merging will typically occur in a time span of a second and will 
accelerate debris to very high velocities reaching almost the speed of light. 
The quiescent emission of the quasar nucleus is then up-scattered to the 
range of gamma-rays by this accelerated hot material (debris), i.e. the
spectrum is shifted by $\Gamma^2$ in photon energy and the flux is
boosted by $\Gamma^4$.
The GRB spectra and their light curves are therefore explained by 
the relative orientation between the observer, the beam direction of the 
emission of the active nucleus, and the path of the accelerated material.
This scenario predicts that the spectra and variability of the GRBs would 
be correlated with that of the host AGN.

\subsubsection{Steady gamma-ray emission from quasars}

Already the HEAO-1 observations (Rothschild et al.\ 1983)   
revealed AGN to show emission up to the 100\,keV energy range  
(mainly from Seyfert 1s).
Emission up to 100\,MeV was detected from radio-loud quasars and 
BL Lac objects, for which synchrotron emission is the favoured mechanism.
However, no emission above 500\,keV was detected from radio-quiet AGN 
or QSOs with the three CGRO instruments BATSE, COMPTEL and OSSE
(Sch{\"o}nfelder 1994).
It appears that two different mechanisms are at 
work in the high-energy band depending on the radio-loudness of the objects.
The existence of two mechanisms is supported by the well-established 
difference between radio-loud and radio-quiet QSOs seen in X-rays 
(cf e.g.\ Wilkes \& Elvis 1987; Schartel et al.\ 1996).
Our result fits into this phenomenological picture, because
we found GRB counterparts only for radio-quiet quasars. 
On the other hand, the maximum energy of photons from GRBs reaches up to
several GeV and thus is significantly higher than 
the observed steady 500\,keV emission from radio-quiet AGN and QSOs.
Thus, there is a clear difference in the gamma-ray spectra of GRBs and the 
steady emission of radio-quiet quasars.

Possibly, the MISO and Ariel 5 observations of NGC 4151 
(Perotti et al.\ 1981), a radio-quiet 
Seyfert galaxy, may give some clues the understanding of our results. 
The flux above $\sim$500\,keV from NGC~4151 was found to be
variable by a factor of 3 to 10 on a time scale up to 1 year
as measured with Ariel 5 in  December 1976, and MISO in May 1977
and September 1979. The  observed  
$\gamma$-ray fluxes are  more than an order of magnitude 
higher than the limits reported from the OSSE/COMPTEL monitoring of 
NGC~4151 (Sch\"onfelder 1994) or any previous measurement.
Therefore, such a possible gamma-ray high state seems to be extremely rare.
The variability by a factor of 3 to 10 in the $>$500\,keV range 
on a time scale of less than one year should be compared
with a hard X-ray variability of only up to a factor of two 
(Perotti et al.\ 1981).   

\subsubsection{The model of Leiter (1980)}

One model to explain the variability found at the highest measured
energies was suggested by Leiter (1980).
If a massive (M $> 10^8 M_{ \odot }$) Kerr black hole
exists in the nucleus of NGC~4151, then the Penrose 
Compton-scattering mechanism can lead to the production of
$\gamma$-rays with a cutoff at an energy of about 3 MeV.
The burst is triggered by sporadic injection of hot plasma
caused by turbulence in the inner region of the accretion disk.
The resulting $\gamma$-ray burst is emitted within an angle
of 40\degr\, from the equatorial plane of the Kerr black hole.
The duration of such an event is given by the 
size of the target region divided by the speed of light which
is located between the event horizon and the ergosphere. 
In most unified models it is usually assumed that radio-loud quasars 
contain a rotating black hole matching the criteria of  Leiter's simulations. 
But for radio-quiet quasars black holes without rotation
are generally favoured. However, the accretion of matter through an accretion 
disk permanently transfers angular momentum to the black hole. 
The increasing angular momentum causes an increase of the
ergosphere. Finally, the ergosphere is larger than the event horizon
creating a very small target region for Penrose Compton-scattering.
Further investigations have strengthened the possibility to produce
GRBs in hot tori around black holes (Jaroszynski 1996).

If we assume that at least a part of the optical luminosity of the quasar
is caused by turbulence in the accretion disk, then
the positional coincidence we found between GRB and intrinsically 
{\it bright} quasars can be explained with a higher injection rate 
of hot plasma and a higher transfer of angular momentum
according to Leiter's model.
Since particularly  the X-ray emission of quasars is interpreted
as a measure of the activity of  the inner part of the
accretion disk, a search for positional coincidences between 
{\it X-ray bright}, radio-quiet quasars and GRBs would be most promising.
It is interesting that the model proposed by  Leiter already 
predicts a focusing orthogonal to the angular momentum of the
black hole. This may explain our negative result obtained for
radio-loud quasars and BL Lac objects if we assume that the
radio-jet is emitted along the rotation axis.

\section{Concluding remarks}

The aim of our study was to develop a method to handle the large 
error boxes of GRBs and to study the positional coincidences 
with QSOs and AGN. We found the surprising result that the nearest and 
intrinsically brightest radio-quiet quasars have a probability of $>$99.7\%
of being associated to the 134 best localized (and thus in general brightest
and nearest) GRBs of the 3B catalogue. While previous claims of correlations 
have usually been found spurious in light of improved (and larger) data samples,
it is not the fact of the mere correlation which is surprising, but rather 
the high probability we find.

If our results are confirmed in future studies with improved statistics
then bright quasars emit $\gamma$-ray bursts at an emission rate of 
$0.015\pm0.004$ bursts per year and quasar for the BATSE sensitivity limit, 
based on a comparison of the 134 
brightest BATSE bursts with 967 quasars of an ``optimized quasar'' sample. 
In order to determine the energy emitted during a  burst 
we consider only the 20 GRBs with a single quasar in their error box.
For 16 of them the fluxes are given in the 3B catalog, thus yielding  
a mean total emission of $3.1\;10^{51}$ erg for the 20--50\,keV band, 
of $7.0\;10^{51}$ erg in  the 50--100\,keV band, and $3.2\;10^{52}$ erg 
in the 100--300\,keV range, respectively.
An isotropic emission over  $4 \pi$ $sr$ was assumed and the following
parameters were used:
H$_0$=50 km s$^{-1}$\,Mpc$^{-1}$, and $q_0= 0.5$ assuming  power law
spectra with  $\alpha  = 2.0 $.

The deviation of the cumulative peak flux distribution of GRBs from the
--3/2 slope should then be used with caution to study the curvature
of space-time, because quasars themselves  show an intrinsic evolution.

We neither wish to discuss the implications of our results on models, nor to 
build a theoretical scenario of GRB emission from radio-quiet quasars.
However, the brief review of the scenarios proposed by Epstein \etal (1993) 
and Leiter (1980) shows that there are 
already theoretical ideas matching our findings, and (in the case of Leiter's
scenario) that phenomena related 
to neutron stars (or stellar-sized compact objects) are not the only possible 
explanation for GRBs. Also, the measurements of NGC~4151 provide evidence for 
emission from radio-quiet AGN and QSO in the energy range required for GRBs.   
This implies that in the Epstein \etal (1993) scenario the boosting during 
the GRB is rather moderate. Thus, our finding of a positive correlation
of GRBs with radio-quiet AGN combined with some previously suggested
scenarios of cosmological GRBs may provide a new avenue towards understanding
the GRB puzzle.

\begin{acknowledgements}
We thank W. Wamsteker and D. Reimers for fruitful discussions.
We appreciate the comments of two anonymous referees.
JG is supported by the Deutsche Agentur f\"ur
Raumfahrtangelegenheiten (DARA) GmbH under contract No. FKZ 50 OR 9201.
\end{acknowledgements}

\end{document}